# MI 2 MI: Training Dyad with Collaborative Brain-Computer Interface and Cooperative Motor Imagery Tasks for Better BCI Performance

Shiwei Cheng, Jialing Wang

School of Computer Science, Zhejiang University of Technology, Hangzhou, 310023, China

**Abstract**: Collaborative brain-computer interface (cBCI) that conduct motor imagery (MI) among multiple users has the potential not only to improve overall BCI performance by integrating information from multiple users, but also to leverage individuals' performance in decision-making or control. However, existed research mostly focused on the brain signals changes through a single user, not noticing the possible interaction between users during the collaboration. In this work, we utilized cBCI and designed a cooperative four-classes MI task to train the dyad. A humanoid robot would stimulate the dyad to conduct both left/right hand and tongue/foot MI. Single user was asked to conduct single MI task before and after the cooperative MI task. The experiment results showed that our training could activate better performance (e.g., high quality of EEG /MI classification accuracy) for the single user than single MI task, and the single user also obtained better single MI performance after cooperative MI training.

## 1 INTRODUCTION

Collaborative brain-computer interface (cBCI) consists of two categories [1], centralized pattern and distributed pattern. The former one collects EEG information of multiple users synchronously, decoding them with a data server to offer final prediction results. For the latter one, each user's EEG data will be processed individually and independently via a subsystem. And the collaborative data are analyzed by an ensemble classifier via voting system. The main achievement of the cBCI system is that it outperforms the single-user BCI systems.

Motor imagery (MI) is one of the conventional BCI paradigm, which has been studied in the past few decades and has strong potentials in rehabilitation and recovery field. In order to improve the performance of the MI training, researchers proposed an approach of MI-cBCI. In this case, instead of single user, a dyad are asked to finish specific MI task for different purposes, such as competition and cooperation [2]. These studies attempts to analyze user's EEG attributes by, for example, power spectrum density (PSD) [3] in various time periods and frequency bands. The analysis of these work usually conclude the changes in the neural activities of single subject in each dyad, which overlooked the interaction effect between the users in the dyad.

However, some interactive elements in the multi-people interactive environment, such as bidirectional and dynamic, are different from single user environments. And these attributes of brain signals, from our knowledge, has yet to be investigated in MI-cBCI tasks. In order to study the way that different brain regions communicate with each other in each dyad, hyperscanning has been used to track the simultaneous activity of human brains [4]. There are several neuroimaging techniques that can be used for hyperscanning, such as functional Magnetic Resonance Imaging (fMRI), electroencephalogram (EEG) and functional near-infrared spectroscopy (fNIRS). Since EEG has the highest temporal resolution and can be used in various scenarios, in this study we use EEG as the tool of hyperscanning to record the dyad's neural activities.

In this study, to evaluate how would the neural activities be influenced through the multi-people scenario in MI task, we proposed a cooperative four-classes MI task that involves 10 dyads to finish the task. And single MI task is also designed for further comparison. Inspired by the existed research work of Cheng et al. [5], in order to guide our subject to better conduct MI task with high-quality EEG signals, we introduced the humanoid robot to posture the guidance stimulation. In addition, we used eye tracking to investigate subjects' recognition state during the whole experiment to ensure their focus on the robot [6]. By applying the hyperscanning technique, we analyzed the neural activities couplings between each dyad in cooperative MI task. Furthermore, we analyzed the neural activities changes from each participant in single MI task. The results suggested that such cooperative MI task could activate better performance for each single user than single MI task, and single user also obtain better single MI performance after dyad based cooperative MI training.

## 2 RELATED WORK

### 2.1 MI-cBCI

MI refers to imagining body movements without actual posturing [7]. Most MI-cBCI paradigms attempts to evaluate the changes of user's neural activities in either cooperative or competitive tasks. For example, Daeglau et al., designed a competitive multi-user race condition using humanoid robots [2]. The robots will have a walking competition based on users' ERD amplitudes in the continuous MI tasks. Bonnet et al., designed a multiuser BCI Game based on MI named BrainArena, which is a football game controlled by hand MI. This game has single, collaborative and competitive tasks, and the latter two tasks involve two users [8]. According to user's classification accuracy result of the MI tasks, the direction of force will be presented. In the collaborative task, the force will be summed from the dyad, while in the competitive task the force will be counteracted. Apart from these collaborative or competitive forms in MI tasks, there are several researches ask the multi-user to conduct the same MI tasks simultaneously without any interactions. For example, Gu et al., proposed an optimization of task allocation for MI-cBCI, they suggested that users should be assigned different MI tasks while maintaining the total MI tasks unchanged. They compared the classification accuracy from such division-of-work and common-work strategy's and the results suggested that by division-of-work could effectively enhance the cBCI classification performance and reduce the individual workload. Zhou et al., explored the MI-cBCI performance in idle detection and compared it with the single task. And the results proved that the multi-user task can perform better than single task by fusing the common spatial pattern (CSP) features of each dyad. All these pioneer work demonstrated that the multi-user scenarios have great potentials to improve the MI performance than single MI task.

### 2.2 Hyperscanning

There is a noticeable problem that aforementioned MI-cBCI work have in common: they overlooked the interactions between the dyad or the group. Apart from the single task, the present of the partner or opponent can have an impact on the user in either positive or negative way, which can be reflected on the neural activity changes. From the traditional cBCI studies, hyperscanning is often introduced as an approach to track the simultaneous changes of neural activities in multi-brains during the various interaction categories, such as competition, cooperation, communication.

In most hyperscanning-related studies, the inter-brain synchronization (IBS) [9] of the dyad could always be investigated. IBS is the key to information exchange between neurons in different regions of the brain, and reflects the type of neural activity that the brains are undergoing during the interaction. IBS has been applied in various competitive and cooperative themes, such as P300 [10], selective attention task [11], art [12], VR [13], and so on. Phase locking value (PLV) [14] is one of the most common-used measurement to evaluate the IBS, it can accurately reflect the changes in the brain network caused in different time and frequency domains, revealing how information is integrated and disseminated in various regions of the brains, and helping to deepen the understanding of the interaction process. In addition, it is reported that IBS can represent how much do the users from the dyad is willing to cooperate with each other in the cooperative task [15]. However, IBS has been scarcely discussed in the MI-cBCI studies, which leaves us much room to understand the coupling of neural activities among various brain regions. Although it has been suggested that the multi-subject environment would have certain influence to the participants in MI task [8], it remains unknown whether such influence is positive or negative. We believe it will be effective and beneficial to calculate IBS and investigate the coupling neural activities of cortical regions in interactive MI scenarios.

## 3  MI-CBCI EXPERIMENT

### 3.1  Task design

In this study, we designed a robot-guided MI task in a cooperative task, in which a humanoid robot will be used as the stimulation display to perform specific postures and guide our dyads to conduct corresponding MI tasks. In addition, a set of MI tasks in single task (left-hand versus right-hand, tongue versus foot) are also applied before and after the cooperative MI task for further comparison. In order to naturally guide the participants, in left/right-hand MI task, the robot will present a left/ right-handshaking gesture, so the participant only need to imagine reaching out the same hand. In the tongue/foot MI task, in order to avoid distracting visual attention, we let the robot present nodding or shaking head twice in each trial, respectively, so that our participant only needs to focus on the robot head. The definition of each task is as follows:

*Single task*: two participants in each dyad are assigned randomly with either left/right-hand MI task or tongue/foot MI task. Assuming participant in a dyad are A1 and A2, respectively, and A1 is asked to perform the left/right-hand MI task according to the gesture, then A2 needs to finish the tongue/foot MI task. Each participant will perform the single mode twice, that is, before and after the cooperative task. And in each single task, their task won't be altered. Note that we denoted the single task before cooperative task as "Phase 1", and that after cooperative task as "Phase 3".

*Cooperative task*: This task requires each dyad to work together to accomplish the task. The robot will randomly perform a combined postures from left/right-hand and tongue/foot MI tasks. For example, the robot reaches out its left-hand while nodding, suggesting A1 should conduct left-hand MI while A2 should conduct tongue MI simultaneously. In this task, we did not force our participant to perform the corresponding MI as they've experienced in the single task, which indicated that they could freely assign their task. Each dyad is allowed to communicate verbally without any body movement. Note that we denoted the cooperative task as "Phase 2".

The specific paradigm design is shown in Figure 1. There are three periods of each trial of MI task: (1) The robot maintains an idle state for 3 seconds, and the participant can take a short rest during the 3 seconds; (2) The robot plays the prompt tone of "Ready" for 1 second, indicating that the participant is about to start the trial; (3) The robot will randomly perform the guidance gesture in 6 seconds, the participant (in single task) or the dyad (in cooperative task) should conduct corresponding MI during this period. In the whole experiment, each participant needs to complete 6 blocks of single task (3 blocks before and after the cooperative task) and 3 blocks of cooperative task, each block includes 20 trials. After the completion of each block, participants would have 3 minutes for relaxation, so as to avoid noise data caused by fatigue. The EEG cap recorded the user's EEG signal during the whole experiment, and automatically labeled different visual stimuli, which was convenient for subsequent classifier training. The eye tracker recorded information about subjects' pupils to ensure their focus on the robot. The total experiment time for each subject was about 60 minutes. At the end of the experiment, an interview was conducted to collect the subjective feedback from all the participants.

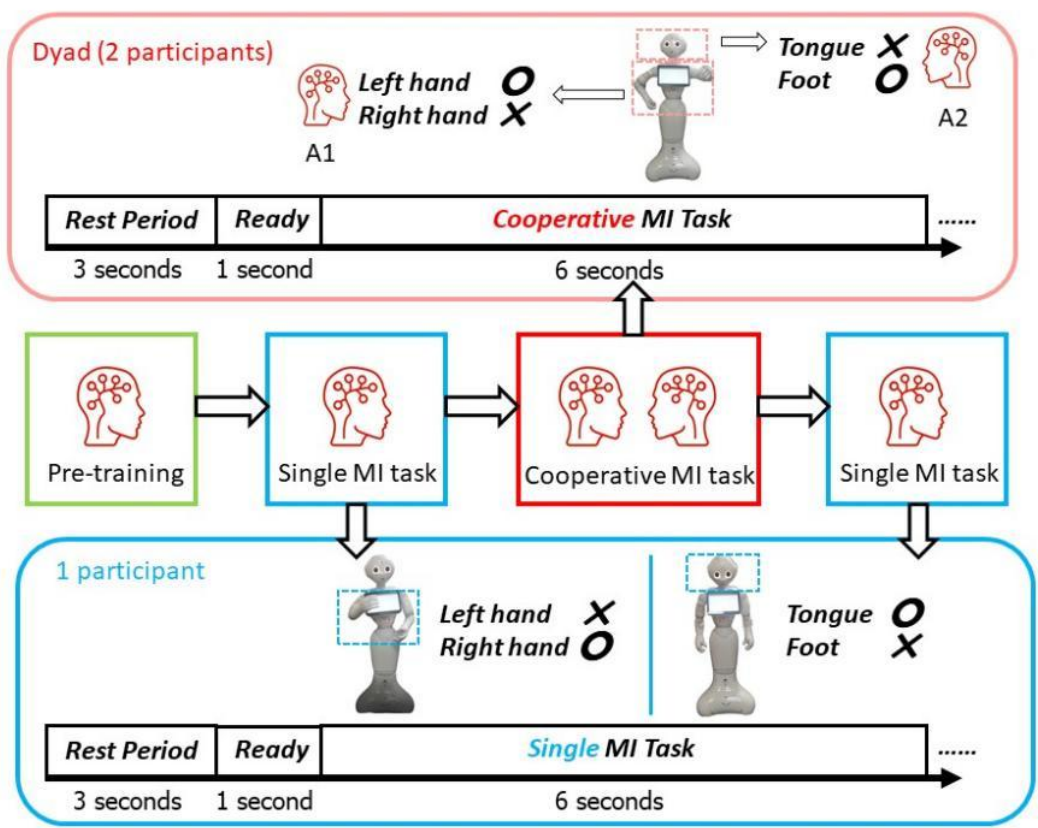

Figure 1: Paradigm of the experiment.

The experiment was conducted in a closed and quiet room. The participants were asked to sit in front of a monitor at a distance of about 100 cm. The experimenter helped the participants to wear the EEG cap and played introduction video of MI on the monitor, and guided the participants how to perform MI. After the participants confirmed that they could independently perform MI, we conducted 40 trials of pre-experiment to acquire their EEG data for off-line model training. Then the experiment officially began, and the EEG cap would record the EEG data of the subjects throughout the whole process. The experiment scene of single task and cooperative task are shown in Figure 2a and 2b.

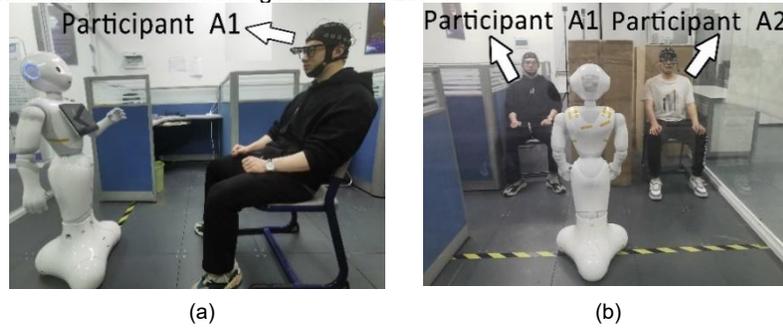

(a)                (b)

Figure 2. Experiment scenario (a)Single MI task; (b) Cooperative MI task.

### 3.2 Participants

Pioneer work has revealed that, during the MI task, people with impaired limbs can have motor cortex activation which is similar to the healthy people [16]. So, to easily recruit enough participants, most related work recruited healthy people in MI study [17]. In this context, our experiment recruited 20 healthy participants without limb disability (8 females and 12 males with an average age of 24.8 years). All participants were proved to be right-handed through the Edinburgh Handedness Questionnaire [18] and passed the Ishihara color blindness test [19], and their vision was in the normal range or had been corrected to normal. This was the first time for all participants to participate in BCI experiment. Before the experiment began, all participants signed informed consent forms and were told the general procedure of the experiment. All the subjects were evenly divided into two groups of four women and six men, and then randomly paired to form 10 dyads. All experiments were approved by the laboratory ethical committee of the authors' university.

### 3.3 Architecture of MI-cBCI

The architecture of the MI-cBCI is shown in Figure 3. The entire MI-cBCI is developed based on OpenCV2 and PySide2. There are two main modules: data processing module and interaction module. The data processing module consists of three sub-modules: data acquisition, data processing and data reservation, and the interaction module consists of two sub-modules: information feedback and data synchronization.

The detailed process of cooperative task is as follows: First, the participant's MI-EEG data is collected in the pre-training phase to train and save the off-line model. Second, the cooperative task begins, and the dyad's EEG data will be collected and sent to the pre-trained model in cooperative task as to conduct real-time classification, and the classification results are transmitted to the interaction module. Meanwhile, interaction module also saves the collected data for further data analysis. Third, the interaction module synchronizes the dyad's EEG data through Transmission Control Protocol (TCP) [20], dynamically feeds back the results to the dyad and the experimenter; Finally, after accomplishing the cooperative tasks, the dyad can make adjustments according to the feedback result and their own strategies, so that they might perform better in the next block of experiment. The process of single task is similar to the cooperative task.

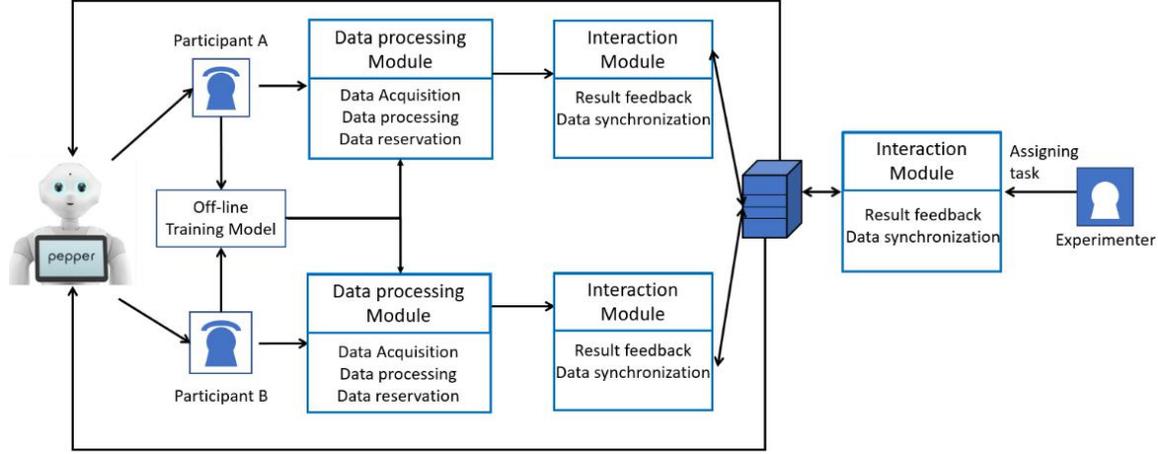

Figure 3. The architecture of the MI-cBCI.

### 3.4 Analysis methods

In this study, we used Neusen W 8-electrode EEG cap to collect the EEG data with the sampling rate of 1000Hz. Since the electrodes C3 and C4 contains the distinct EEG features of imagining hand movement, and the electrode Cz contains the distinct EEG features of imagining the feet and tongue, we selected these 3 electrodes in data acquisition. In addition, in order to analyze the flow of information (the direction of synchronization between brain and brain) during collaborative tasks among a group of subjects, this study set up three regions of Interest (ROI). They were parietal lobe (P3, Pz, P4 electrodes), prefrontal lobe (Fp1, Fp2 electrodes) and central region (C3, C4, Cz electrodes), hence there is a total of 8 electrodes.

#### 3.4.1 Inter-brain synchronization

In this study, we use phase locking value to calculate the IBS between each dyad in the cooperative task. PLV is a statistical measure that can well suit for capturing the rapid flow of information between people through their interaction. The calculation formula is as follows:

$$PLV = \frac{1}{N}\left|\sum_{n=1}^{N} e^{(J(\Phi_x(t)-\Phi_y(t)))}\right|$$

where $N$ represents the number of trials, $\Phi(t)$ is the phase, and $x$, $y$ are the electrode from participant A1 and participant A2 in a dyad, respectively. The calculation of PLV usually obtained after summing and averaging the instantaneous phase difference of multiple trials. The PLV ranges from "0" to "1", where "1" indicates that the phase between the two electrodes from the dyad is completely synchronized, and "0" indicates that the phase between two electrodes from the dyad is totally unsynchronized.

Since the IBS in MI-related task has yet to be discovered, we used band filters to obtain 5 common-used frequency band, that is, δ (0.5-4Hz), θ(4-8Hz), α(8-13Hz),β(13-30Hz) and γ(30-60Hz). A notch filter was also used to remove the 50Hz interference in authors' country. And the EEG data was then down-sampled to 250Hz. Then, the EEG data was divided into several data segments with a length of 10 seconds (the same as a single trial duration). To make this study more convincing, all data were examined before and after preprocessing to ensure that all necessary EEG data were collected and that there were no invalid or missing data segments. The IBS of the rest state (0-4 seconds) and task state (5-10 seconds) were both calculated in 5 frequency bands across the 60 cooperative trials. Both the sliding window and the step length was set as 100 ms.

### 3.4.2 Functional brain network

Furthermore, we also use PLV to investigate the connectivity between the regions of single user's brain, and construct the functional brain network as it may In this study, we attempt to explain the user's MI performance by comparing the brain network structures before and after the cooperative task. In order to study the differences of participant's neural activities before and after cooperative task, we introduced the functional brain network (FBN) and analyzed different measures in the FBN because it could reveal valuable information on how different brain regions communicate in the single MI task [21]. According to the graph theory, the whole brain can be regarded as a complex network, the electrodes can be seen as "nodes", and the connectivity between each pair of electrodes can be seen as "edges". According to these basic elements, the FBN of each participant can be constructed.

To characterize the organization of the FBN, we selected the following metrics [19]: (1) *characteristic path length*, which reflects how fast the global information propagation is; (2) *clustering coefficient*, which reflects how much does the local network efficiency increase; (3) *small-worldness*, which is determined by the characteristic path length and clustering coefficient; (4) *degree centrality*, which reflects the importance of nodes in the whole network and the information communication capabilities; (5) *betweenness centrality*, which is defined as the number of shortest paths going through a node or edge.

The main steps of constructing FBN is organized and depicted in Figure 4. The pre-processing is same as in IBS. Note that we discarded the rest state data, and remained the task state data to calculate the connectivity. The selected 8 electrodes were regarded as "nodes" and we calculated the PLV between each two nodes, and the calculation results, that is, the connectivity between each two nodes, were used as "edges" in the brain network. Since the noise data can also contribute to the weak connection between electrodes, it is necessary to set a threshold for the network to discard spurious connections [23]. Inspired by Bassett's work [24], we selected the threshold as large as possible while guaranteeing that all nodes are connected in five frequency bands. We used the threshold to remove the weak connectivity between the nodes, and we used circular graph to generate the visualization of the FBN.

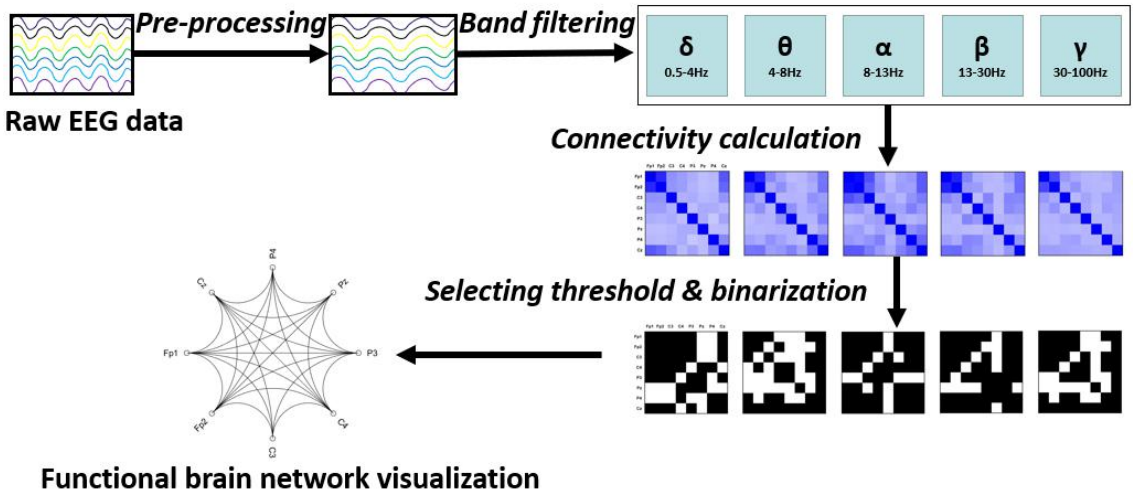

**Raw EEG data**

*Pre-processing*

*Band filtering*

| δ | θ | α | β | γ |
|---|---|---|---|---|
| 0.5-4Hz | 4-8Hz | 8-13Hz | 13-30Hz | 30-100Hz |

*Connectivity calculation*

*Selecting threshold & binarization*

**Functional brain network visualization**

Figure 4. Construction of the functional brain network based on EEG data.

### 3.4.3 Classification accuracy

The MI-EEG data contains much temporal and spatial information, and the classification accuracy can be improved if we take fully advantages of these information. In order to better predict the classification result, we used a combined 2D CNN-LSTM model. This model consists of two modules, the 2D CNN model and the LSTM model. The CNN module is composed of four convolutional layers and two pooling layers. By introducing the wrapper, *Timedistributed*, the temporal information in the EEG data will not be destroyed during the convolution. After extract the spatial features in the CNN module, the features will be sent into the LSTM layer as input, and the dense layer finally output the classification results. The whole framework of the 2D CNN-LSTM model is shown in Figure 5.

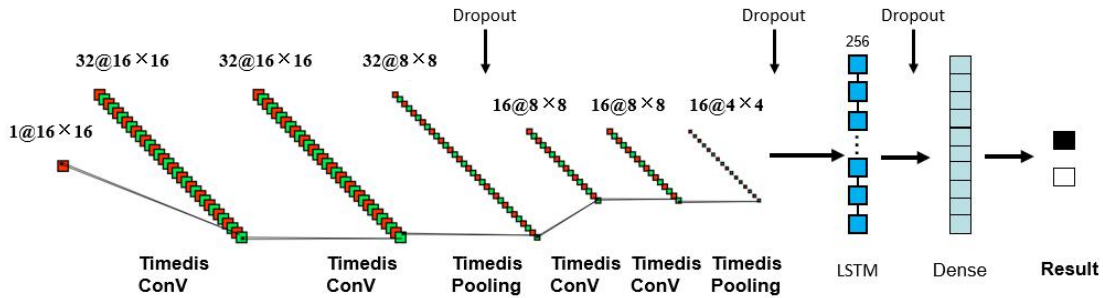

Figure 5. The framework of the proposed 2D CNN-LSTM model.

We used C3, C4 and Cz as they have the most informative features in MI task. The pre-processing is similar to that in calculating IBS, except that only the EEG data in α and β-bands were filtered for feature extraction. In addition, independent component analysis (ICA) [25] was used to remove the noises data. We performed the 10-fold cross-validation over the EEG data from these 3 electrodes via the 2D CNN-LSTM model, and the data was trained for 100 epochs, the batch size was set to 128, and the initial learning rate was set to 0.001. After every 10 epochs, the learning rate is reduced by 20% to avoid overfitting.

## 4  EXPERIMENT RESULT AND ANALYSIS

### 4.1  Inter-brain synchronization

In this study, the IBS of a dyad in four kinds of cooperative tasks (left-hand and tongue, left-hand and foot, right-hand and tongue, right-hand and foot) was calculated respectively in the rest state and task state, as shown in Figure 6, in which the red and black dots are the average IBS of the task state and rest state, respectively. The red solid line and black dashed line represent the change of inter-brain synchronization in the time dimension of task state and rest state, respectively. The average IBS from the task state of EEG signals in $\theta$ and $\alpha$-bands were generally higher than that in rest state. While in the remaining 3 bands, the average IBS from the rest state were higher than that in the task state. The PLV of the $\delta$-band from the rest state and the $\alpha$-band from the task state fluctuated around 0.95, which were the highest PLV of all the frequency bands. This is due to that, during the rest phase, both of the participants were relaxed, therefore, the most obvious $\delta$-band EEG signal with synchronous rhythm could be stimulated. However, during the task state, which requires full concentration, the dyad was cooperating continuously, such behaviors would trigger the $\alpha$-band EEG signal with the feature of synchronous rhythm along with the visual stimulus, making the IBS between the two bands the highest. The temporal distribution of PLV in $\delta$ and $\theta$-bands are sparser than that in the rest three bands, regardless of the states. According to the fitted lines, during the task state, except for the obvious negative slope from the $\delta$-band in the right-hand and tongue task, the fitting slope is basically positive, which demonstrated that each dyad could continue to maintain high IBS during the task state by communication in a cooperative way.

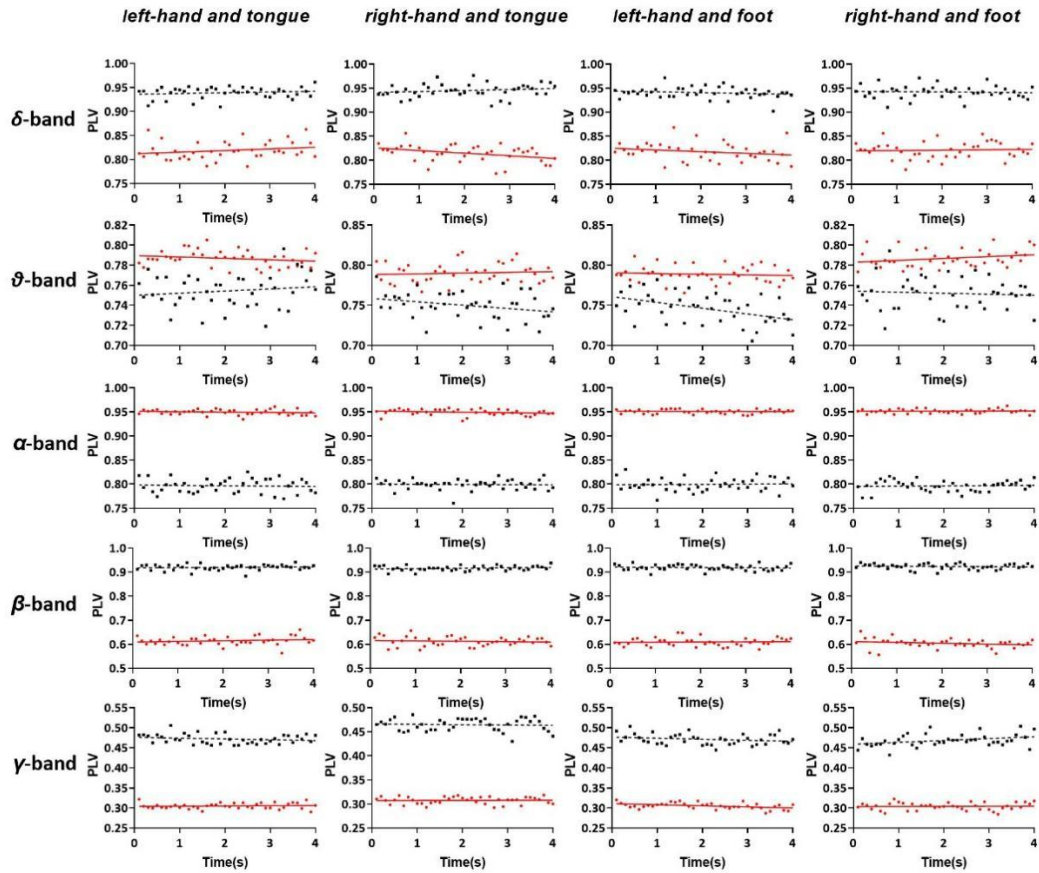

Figure 6. Average IBS distribution of the task and rest states in four cooperative tasks through five frequency bands.

Subsequently, to see if there is a significant difference in IBS between the task state and the rest state over the 60 trials, a paired T-test was carried out, as shown in figure 7. It could be observed that, only in the right-hand and foot task, there was no significant difference between the electrodes in any pairs of electrodes in γ-band, the other tasks contained more than 2 pairs of electrodes in each frequency band, which showed significant difference between the IBS of the task state and the rest state. The number of electrode pairs induced by tongue motion imagination task was significantly higher in δ, θ, β and γ bands than that induced by foot motion imagination task, but showed the opposite trend in α band.

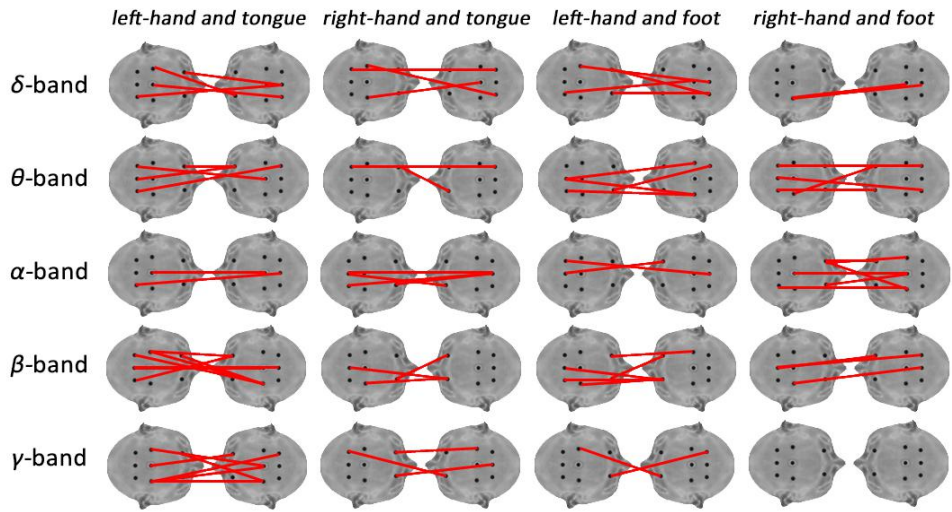

Figure 7. Significant differences of IBS in four cooperative tasks through five frequency bands.

Figure 8 shows the distribution of the total number of electrode pairs with significant differences in each frequency band among different brain regions in four cooperative tasks. The IBS of frontal–central, frontal–parietal and parietal-central regions were significantly improved during the task phase. In addition, imagining the tongue-related cooperative task could activate more significant changes in IBS than that of the foot-related cooperative task. In addition, in the right-hand and tongue task, no significant difference was observed between the central-central region of each dyad.

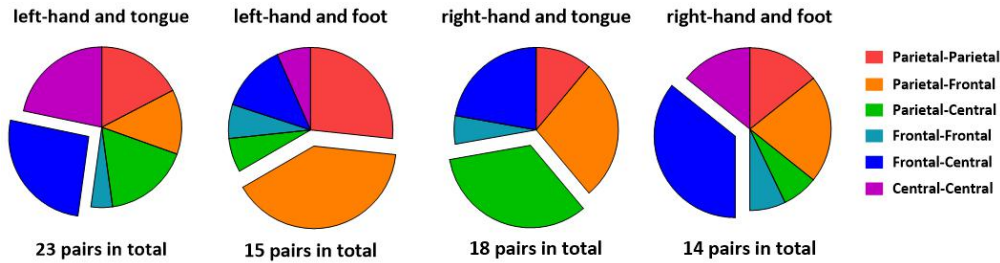

Figure 8. Distribution of IBS with statistically significant difference in all region of interests.

## 4.2 Functional brain network

The functional connectivity of 10 dyads in five EEG frequency bands during the task was calculated by PLV, and then the FBN was constructed with a threshold value of 0.3 to ensure that all the nodes were connected in the network. The EEG data before and after the collaborative task were compared and analyzed in each frequency band across two binary classification tasks, and the paired T-test as used for each node pair to detect whether there was a significant difference in the functional connectivity before and after the cooperative task, the result was shown in Figure 9. Of the five frequency bands, the γ band has the most pairs of electrode pairs with significant differences, and the δ band has the least pairs. In addition, no significant differences were observed between the parietal lobe regions (P4-Pz, P3-Pz) in the α and β bands, only between P3-P4 in several cases. In the left-hand task, the left frontal region (Fp1) and the left central region (C3) always maintained a significant increase in functional connectivity in each frequency band. Similarly, in tongue and foot task, the right frontal lobe (Fp2) and the central region (Cz) also maintained a significant increase in functional connectivity in each frequency band.

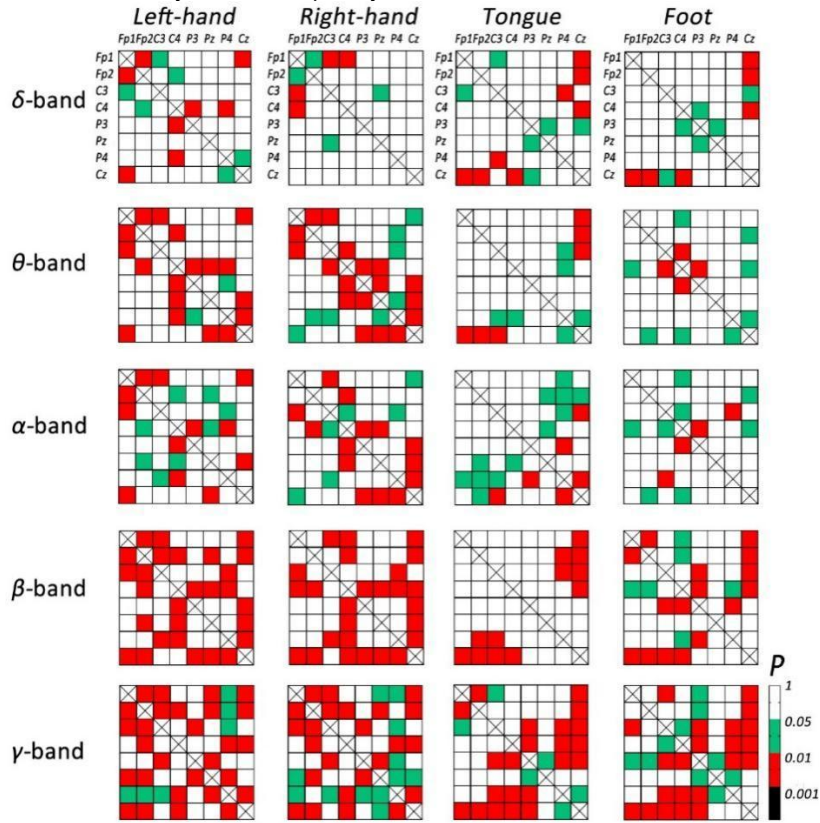

Figure 9. Distribution of the significant differences of functional connectivity in different frequency bands.

We set the threshold as 0.3 and built the FBN in all frequency bands. We used the five metrics: *characteristic path length*, *clustering coefficient*, *small-worldness*, *degree centrality*, and *betweenness centrality*, and presented the comparison results between Phase 1 and Phase 3 in γ band, as shown in Table 1. Paired *T*-test was used to observe whether there was significant difference in these metrics between Phase 1 and Phase 3. It can be seen from the results that after cooperative task, the characteristic path length decreased significantly in the left-hand, tongue and foot tasks, but no significant difference was observed in right-hand task as it only decreased from 2.470 to 2.405 (*p*=0.129). The clustering coefficient was generally improved after cooperative task, and significant differences between Phase 1 and Phase 3 were observed in both left-hand and foot MI tasks (*p*<0.05), but we did not observe the significant differences in the right-hand (*p*=0.385) and tongue (*p*=0.671) tasks. Similar to the clustering coefficient, the small-world attribute showed an increase trend in all single task. Among them, the value increased significantly in the left-hand, right-hand and tongue MI tasks after cooperative task ($p<0.05$), but did not increase significantly in the foot MI task ($p=0.197$).

Table 1: Comparison results of functional brain network of Phase 1 and Phase 3.

| Metric | Task (Phase 1- Phase 3) | | | |
| --- | --- | --- | --- | --- |
| | Left-hand | Right-hand | Tongue | Foot |
| Characteristic path length | 0.086, $p<0.05$ | 0.065, $p=0.129$ | 0.104, $p<0.05$ | 0.378, $p<0.05$ |
| Clustering coefficient | -0.127, $p<0.05$ | -0.030, $p=0.385$ | -0.024, $p=0.671$ | -0.080, $p<0.05$ |
| Small-worldness | -1.935, $p<0.05$ | -2.730, $p<0.05$ | -1.417, $p<0.05$ | -0.872, $p=0.197$ |

The comparison result of degree centrality was shown in Figure 10. In the left-hand task, except for Fp2 and C4, the degree centrality of remaining 6 electrodes was improved after cooperative task, and the degree centrality of C3, P3 and P4 electrodes was significantly improved ($p<0.05$). In right-hand task, only Fp2 showed a significant decrease ($p<0.05$), while the remaining electrodes showed an increasing trend, among which C3 was the most significant, and the average degree centrality increased to 4.28. In the tongue task, the degree centrality of all electrodes was improved after cooperative training, and significant improvement was observed in Fp1, and the central region (C3, C4 and Cz). In the foot task, except for P3 and Pz, the degree centrality of the remaining 6 electrodes was significantly improved after cooperative task ($p<0.05$). These results demonstrated that all these electrodes have gradually shown its potential importance in constructing the brain network after cooperative task.

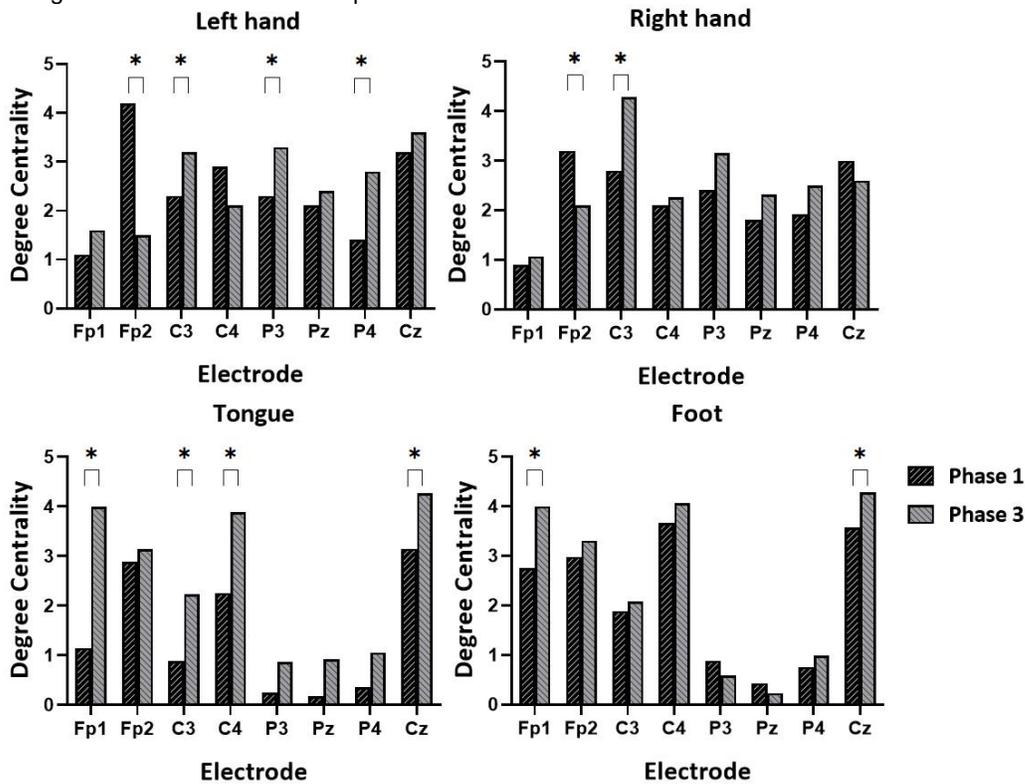

Figure 10. Comparison result of degree centrality of Phase 1 and Phase 3.

As shown in Figure 11, in the left-hand task, the betweenness centrality of Cz showed a decrease ($p$=0.791) in Phase 3, the remaining 7 electrodes showed an increase trend instead, and significant differences were observed at Fp2, C4, P3 and Pz ($p$<0.05). In the right-hand task, except C4 which showed an insignificant decrease ($p$=0.081), the remaining electrodes all showed an increase tread, and a significant difference was observed at C3 ($p$<0.05). In the tongue task, except Pz showed an insignificant decrease ($p$=0.097), the remaining 7 electrodes had higher degree centrality in Phase 3 than in Phase 1, but no significant difference was observed at any electrode. In the foot task, except Fp1 and Fp2, the betweenness centrality in the remaining 6 electrodes had increased, and a significant difference was observed at Cz ($p$<0.05).

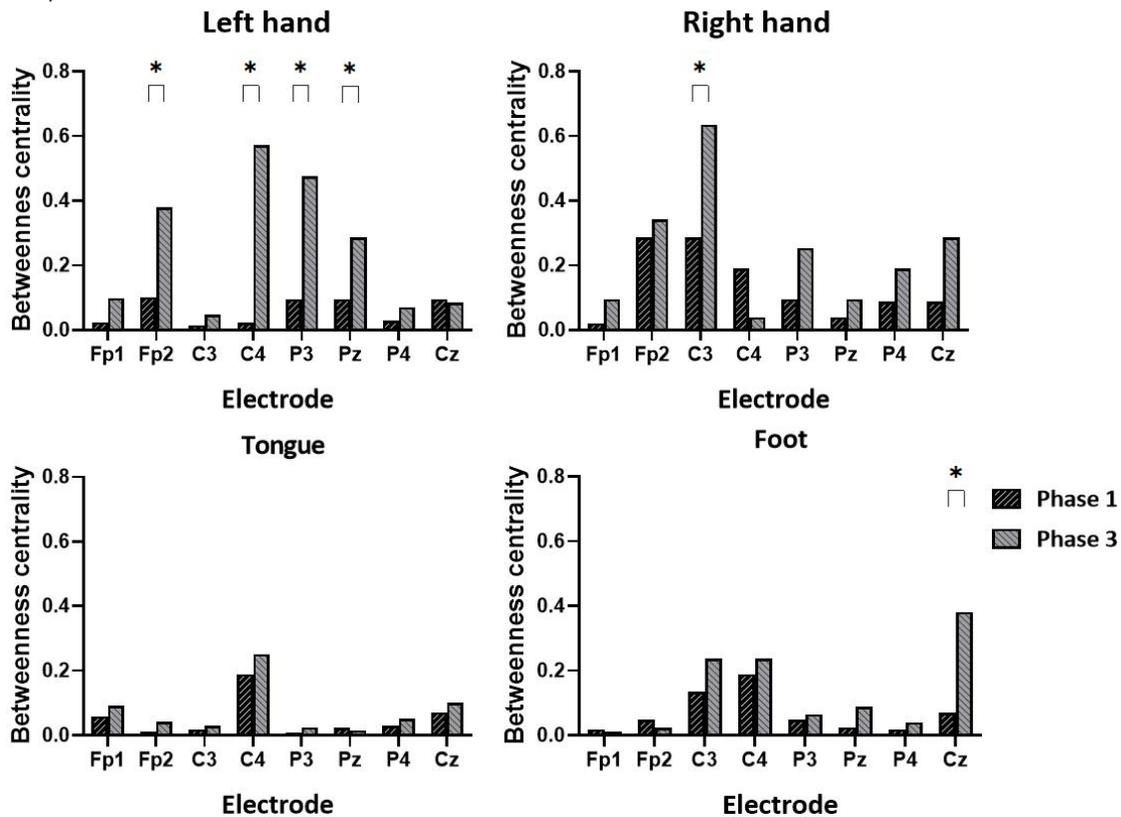

Figure 11. Comparison result of betweenness centrality of Phase 1 and Phase 3.

Based on the threshold value of 0.3, we visualized the FBN via circular graph from MATLAB platform, the FBN in Phase 1 and Phase 3 was shown in Table 2. The results showed that FBN connectivity in tongue and foot tasks was closer than that in left and right-hand tasks. After cooperative training, additional connections were constructed between frontal region and central region in both left and right-hand task. As for tongue and foot task, the parietal region had more strong connection with the frontal region and central region in Phase 3.

These results reflected that the cooperative task might improve the information transfer efficiency of the FBN, and stabilize the topological structure of the FBN.

Table 2: Functional brain network visualization of Phase 1 and Phase 3 with the threshold of 0.3.

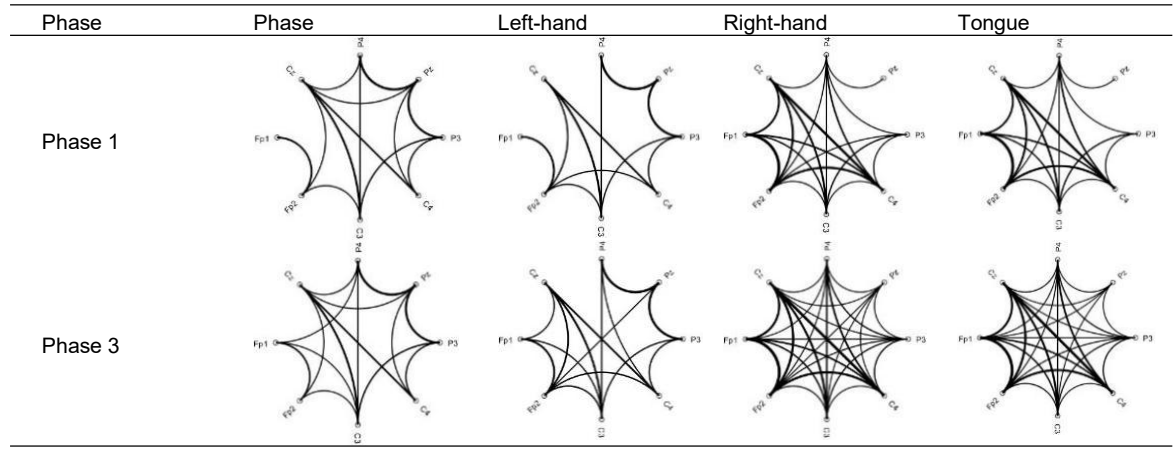

| Phase | Phase | Left-hand | Right-hand | Tongue |
|-------|-------|-----------|------------|--------|
| Phase 1 | | | | |
| Phase 3 | | | | |

## 4.3 Classification accuracy

Figure 12 shows the classification accuracy results of the left and right-hand MI in all three phases. The average classification accuracy in Phase 2 achieved the highest, reaching 90.26%, which was 7.19% and 1.86% higher than that in Phase 1 and Phase 3, respectively. Kruskal-Wallis test was carried out to see if there was significant difference in the accuracy among these 3 phases. The results showed a significant difference of the classification accuracy between Phase 1 and Phase 2 ($p<0.05$), and between Phase 2 and Phase 3 ($p<0.05$), but no significant difference was observed between Phase 1 and Phase 3 ($p=0.858$).

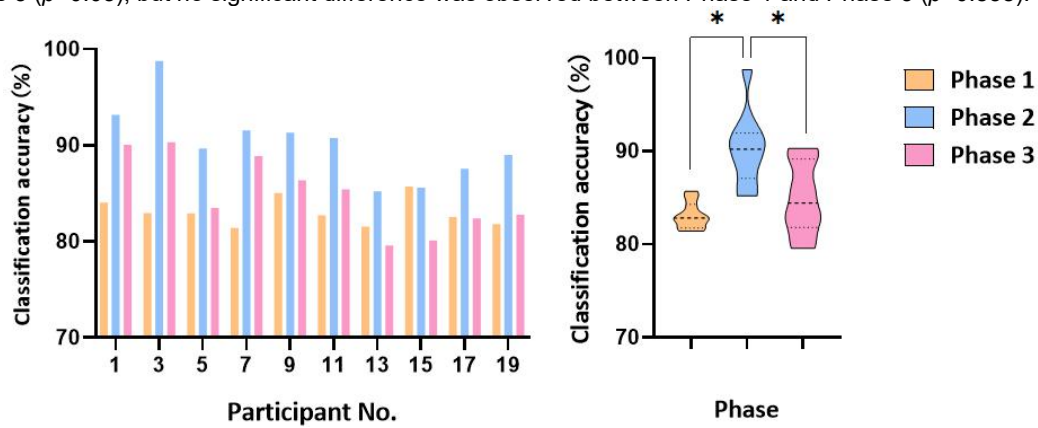

Figure 12. Left- and right-hand MI classification results across all three phases.

Figure 13 shows the classification accuracy results of the tongue and foot MI in all three phases. The average classification accuracy in Phase 2 achieved the highest, reaching 88.78%, which was 8.83% and 3.71% higher than that in Phase 1 and Phase 3, respectively. Kruskal-Wallis test was carried out to see if there was significant difference in the accuracy among these 3 phases. The results showed a significant difference of the classification accuracy between Phase 1 and Phase 2 ($p<0.05$), but no significant difference was observed between Phase 2 and Phase 3 ($p$=0.076), and between Phase 1 and Phase 3 ($p$=0.062).

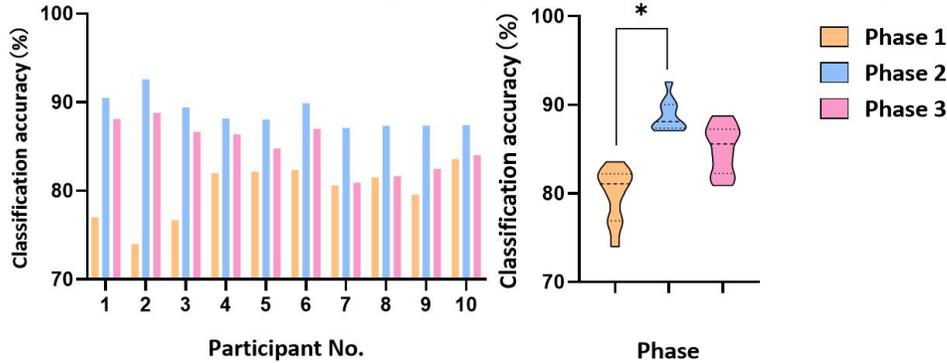

Figure 13. Tongue and foot MI classification results across all three phases.

Table 3 shows the average F1-score of both left and right-hand task and tongue and foot task across all three phases. It could be concluded that in both tasks, Phase 2 reached the highest average F1-score. And Kruskal-Wallis test also proved the significant differences in F1-score between Phase 2 and Phase 1, and between Phase 2 and Phase 3 ($p<0.05$). All these results strongly suggested that the cooperative task could activate best MI performance for the individuals. In addition, if we regard the cooperative task as a training method, it seems this training can assist the individuals to perform better in their own MI tasks.

Table 3: Average F1-score result across all three phases.

| Task | Phase 1 | Phase 2 | Phase 3 | *p*-value |
|------|---------|---------|---------|---------|
| | | Phase | | |
| Left and right-hand | 0.76 | 0.88 | 0.81 | <0.05 |
| Tongue and foot | 0.74 | 0.83 | 0.75 | <0.05 |

## 4.4 Subjective feedback

In this study, we asked all the participants to finish a 5-point Likert scale questionnaire about the MI-cBCI experiment, which was based on the NASA-TLX task load index [26] and the NMM social presence questionnaire [27]:

(1) Mental demand: How much mental activity is needed to imagine the body movement? (1 - easy, 5 - extremely difficult);

(2) Temporal demand: Will you feel nervous during the MI because of the lack of time in completing the task? (1 - never, 5 – very much);

(3) Effort: How much or how little effort did you put into completing the task? (1 points - very few, 5 points – very much);

(4) Presence: Do you feel the presence of your partner during the completion of the cooperative task? (1 score – not at all, 5 score – very much);

(5) Divided attention: Will your attention be divided by the complex body movement the robot presented in the cooperative task? (Score 1 - not at all, 5 – very much).

Figure 14 shows the results of questionnaire for the MI-cBCI experiment. The average score of *mental demand* was 4.6, and all participants scored 4 or 5 points, which meant that all participants needed to fully use their imagination during the experiment. The average score of *temporal demand* was 3.6, and most of the subjects scored 4 points, indicating that the 6-second MI duration was relatively urgent for most users, and there were 3 participants who gave 5 points. One of them pointed out,

> "I was attracted by the appearance of the robot at the beginning, so I did not pay much attention to its actions. When I realized I was in the task, the time was almost up."

The average score of *effort* was 3.7 points, and its distribution was similar to that of temporal demand, but the correlation between temporal demand and effort was not obvious. The average score of *presence* was 3.9, indicating that the participants could basically be aware of the presence of the partner in each dyad. Finally, the average score of *divided attention* was 3.8, indicating that most participants would be distracted by the complex body movement in cooperative task. It was inevitable that some of the participants might imagine their partner's task. Nonetheless, two participants gave 2 points, from the interview, one of them who was responsible for the left and right-hand MI task pointed out,

> "I know he is sitting next to me, but I am not very concerned about whether he has completed his part, I tell myself to focus on my own part, so I didn't look at whether the robot was nodding or shaking head, all my attention is focused on the hand movement. It's a lot easier for me."

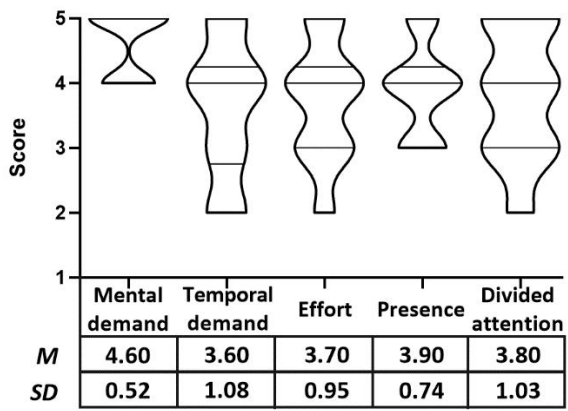

| | Mental demand | Temporal demand | Effort | Presence | Divided attention |
|---|---|---|---|---|---|
| *M* | 4.60 | 3.60 | 3.70 | 3.90 | 3.80 |
| *SD* | 0.52 | 1.08 | 0.95 | 0.74 | 1.03 |

Figure 14. Subjective questionnaire results.

## 5  DISCUSSION

In this study, we analyzed the IBS between 10 dyads in 4 cooperative tasks, and the results showed that the highest IBS (PLV value) appeared in α band, one possible reason is that, during the cooperative task, the dyad can talk to each other, and by discussing and confirming what they've observed from the robot, the α-band wave will provide a state of readiness to perform cognitive activities [28].

Specifically, in left- and right-hand single task, the significant increase of the pairs of electrodes mainly involved the frontal lobe (Fp1, Fp2), we believed this is due to the gesture, that is, hand shaking can be guided, and it showed the wills of communication with our participant, which can lead our participant to react as in the social environment. Similar result is revealed in Balconi's work [29] where they study how would Italian and French gestures affect the EEG connectivity. A fMRI study also found the parietal and frontal regions IBS with hand movements [30]. This might also explain why in the tongue and foot task, the frontal lobe did not have very close connections with other brain regions.

The main deficiency of the study is that participants are asked to watch the head shaking and conduct foot MI, which can cause troubles for them because imagining foot movement has almost nothing to do with head shaking. This task requires our participants to be very responsive and good at imagining. Such design can be very difficult for those BCI illiteracy, and it will be tough to reach training objective. Therefore, in the future, we will explore more guided and meaningful gestures or postures in foot and tongue tasks, as to assist the participants to gain MI capabilities easily.

## 6  IMPLICATION OF DESIGN

In this study, we have proved that cooperative MI training could help user to produce higher-quality MI-EEG signals, which can improve the MI classification accuracy. Based on this finding, conducting multiple user MI can guide people with movement disabilities to keep activate the motor cortex and leverage their motor rehabilitation. Therefore, therapists can assign cooperative MI tasks to people with movement disabilities to better train their MI capabilities. In addition, by analyzing the IBS metric, we could find out who shows a willingness to cooperate or has the MI capabilities to fill the gap for certain brain region dysfunction, so the family members or friends of people with movement disabilities have a chance to be companions for the MI training, and help improve the quality of training. Furthermore, training by cooperative MI task can help people with disabilities facilitate daily life, such as sending control commands more accurately to manipulate the external devices like service robot or powered exoskeleton system. For example, we can use a humanoid service robot to serve them drinking a cup of water. By repeated training in the cooperative task, people with motor disabilities can send the control commands more quickly and precisely.

## 7  CONCLUSION

In this study, we designed a cooperative MI task that required a dyad to accomplish the complex MI task simultaneously. In order to explore whether such cooperative pattern will benefit participants' MI performance, we also designed two phases of single MI task before and after the cooperative task. To evaluate the performance of the dyad, we applied the inter-brain synchronization and functional brain network on the basis of the phase locking value. The information flow and interaction mechanism of EEG signals in different frequency bands in cooperative task were revealed, as well as the characteristics of participant's FBN in single task. The inter-brain synchronization of parietal-central, frontal-central and parietal-frontal regions showed significant improvement in the task state. And the metrics of the functional brain networks were significantly enhanced after collaborative tasks in the single task. The classification accuracy of cooperative task reached 90.26% (left- and right-hand task) and 88.78% (tongue and foot task), which outperform both of the single tasks, nonetheless, after the cooperative task, the performance in the single task yielded better accuracy than that before the cooperative task. As a preliminary study for the MI-cBCI, our study has proved that the cooperative task in motor imagery shows great promises as a guidance training motor imagery capabilities.

## ACKNOWLEDGMENTS

The authors would like to thank all the volunteers who participated in the experiments. This work was supported in part by the Natural Science Foundation of Zhejiang Province under Grant LR22F020003, and the National Natural Science Foundation of China under Grant 62172368.